# Continuously tunable pulsed Ti:Sa laser self-seeded by an extended grating cavity

Ruohong Li,[1,*] Jens Lassen,[1,2,3] Sebastian Rothe,[4,5] Andrea Teigelhöfer,[1,2] Maryam Mostamand,[1,2]

[1] *TRIUMF, 4004 Wesbrook Mall, Vancouver, British Colombia, V6T 2A3, Canada*
[2] *University of Manitoba, Winnipeg, Manitoba, R3T 2N3, Canada*
[3] *Simon Fraser University, Burnaby, British Colombia, V5A 1S6, Canada*
[4] *CERN, Engineering Department, CH-1211 Geneva 23, Switzerland*
[5] *School of Physics and Astronomy, The University of Manchester, Manchester, M13 9PL, UK*

**Abstract:** A continuously tunable titanium:sapphire (Ti:Sa) laser self-seeded by an extended grating cavity was demonstrated and characterized. By inserting a partially reflecting mirror inside the cavity of a classic single-cavity grating laser, two oscillators are created: a broadband power oscillator, and a narrowband oscillator with a prism beam expander and a diffraction grating in Littrow configuration. By coupling the grating cavity oscillation into the power oscillator, a power-enhanced narrow-linewidth laser oscillation is achieved. Compared to the classic grating laser, this simple modification significantly increases the laser output power without considerably broadening the linewidth. With most of the oscillating laser power confined inside the broadband power cavity and lower power incident onto the grating, the new configuration also allows higher pump power, which is typically limited by the thermal deformation of the grating coating at high oscillation power.

*[*ruohong@triumf.ca](*ruohong@triumf.ca)*

**keywords:**  laser cavity, Ti:Sa laser, tunable laser, diffraction grating.

## 1.     Introduction

A nanosecond pulsed Ti:Sa laser with broadly tunable wavelength range, single-longitudinal-mode operation and high peak power, is a powerful tool used for scientific researches and applications, such as resonance laser ionization spectroscopy [1,2], laser isotope separation [3,4], non-linear optical process [5] and Light Detection and Ranging (LiDAR) [6]. Wavelength selection for this type of Ti:Sa laser is typically realized by use of a birefringent filter (BRF) or a diffraction grating. A BRF usually introduces lower cavity loss and therefore provides higher output power as compared to a grating laser. However, to obtain GHz laser linewidth the BRF laser must be augmented with an additional etalon, which makes a continuous wavelength scan beyond 0.2 nm laborious, as both wavelength selective elements must be tuned correlatively. Wavelength tuning of a grating laser is achieved by changing the tuning element angle: the grating angle for Littrow configuration or the feedback mirror for Littman configuration [7]. This allows for smooth and continuous wavelength tuning and therefore has been implemented in most tunable Ti:Sa laser systems employed for spectroscopy.

Due to the limited reflectivity of grating coatings, aluminum or gold, in the Ti:Sa tuning range, grating lasers normally have lower output power compared to BRF laser systems. In addition to the reflection losses, local heating and deformation of the grating at high intra-cavity power also limits the overall output power of grating laser based systems. Various improvements have been attempted to make up for these shortcomings. A common method is to add a power amplifier with active medium using either single or multi-pass amplification [8]. However such design significantly increases the complexity of the laser system at the same time as improving the power output.

To enhance the performance with simple optics, a dual-cavity configuration had been demonstrated by Ko *et al.* through adding a feedback mirror for 1st order diffraction on a standard Littman configuration [9]. With the oscillation higher above the threshold, a further enhanced operation of this configuration was reported later [10]. Afterward a double-grating oscillator was introduced as a modified version by changing the high reflection feedback mirror with a grating [11]. The wavelength restriction of this double-grating geometry narrowed down the laser linewidth to ~ 370 MHz compared to conventional single grating geometry. A combination of the dual-cavity design with the active medium amplification was also experimented [12].

All these demonstrated dual-cavity designs have a shorter narrowband (NB) cavity than the broadband (BB) amplifier cavity to avoid the depletion of inversion population before the NB cavity establishment. However early in 1971, Bjorkholm *et al.* [13] had proposed a simple dual-cavity configuration (called as mirror-grating combination in their paper) with a NB cavity longer than the BB cavity. A successful experiment on the configuration was also briefly demonstrated.

In this work, Bjorkholm's configuration was adapted and further developed. A partial reflection (PR) mirror was inserted inside of the conventional Littrow configuration single-grating Ti:Sa pulsed laser cavity to construct two oscillators. Most irradiation is retro-reflected by the PR mirror to build a BB Fabry-Perot (F-P) resonator. A smaller amount of that is transmitted to the diffraction grating. The grating and the PR mirror constitute a parallel NB oscillator with wavelength selection. By coupling the two cavities together, the grating cavity behaves as an injection seeding laser and the F-P cavity as a slave oscillator or a power amplifier. This parallel dual-cavity configuration lowers the total cavity loss on the grating dispersion without trading off on the narrow linewidth. Furthermore it increases the acceptable pump power by reducing the power deposed on the grating surface, therefore allows for further enhancement on output laser power.

## 2. Experimental setup

The experimental setup was based on a grating-tuned Ti:Sa laser previously laboratory-built by our group [14], which is regularly used for in-source laser resonance ionization spectroscopy of short lived isotopes [15]. The schematic of the setup is shown in Fig. 1. Basically the laser is a typical Z-shaped standing wave resonator. The Ti:Sa crystal is a 6.35 mm diameter and 20 mm long rod doped for 1.9 cm$^{-1}$ absorption at 532 nm. To reduce reflection losses, the two end surfaces are cut and polished at Brewster's angle. The crystal is pumped by a frequency doubled Nd:YAG laser (LEE laser LDP-100MQG, M$^2$=20, 10 kHz repetition rate, 160 ns pulse width), which focuses on the front surface of the Ti:Sa crystal by a planoconvex lens of $f = 75$ mm. A typical pump power is ~ 10 W. The two curved mirrors are mounted at an angle of 18.1° with respect to the laser beam in order to provide for astigmatism compensation. The Ti:Sa resonator mirrors are broadband Ti:Sa laser optics coated for 680-1050 nm.

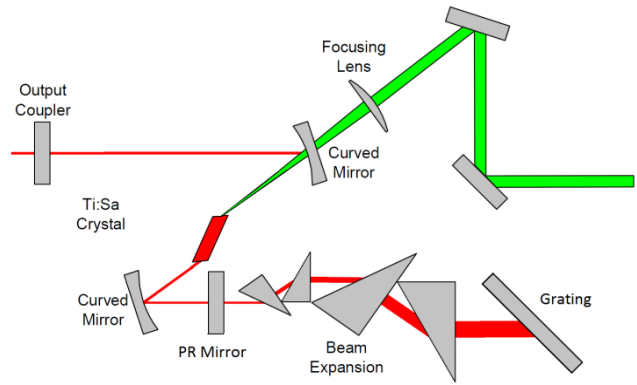

Fig. 1. Schematic diagram of the Ti:Sa laser self-seeded by an extended grating cavity. This is achieved by splitting the classical grating laser resonator by use of a partially reflecting mirror (PR) in the lower parallel arm of the Ti:Sa resonator. Laser performance with different PR mirror reflectivity was investigated.

The grating was purchased from Newport (Richard Gratings$^{TM}$, 53-108ZD02-290R) and has the following specifications: 1800 grooves/mm, gold coated, blazed for 500 nm. It orients in a Littrow configuration with the incident angle rotating from 39.0° to 58.7° corresponding to the wavelength range from 700 nm to 950 nm. Four flint glass (SF11, refractive index = 1.76) 30°/60°/90° prisms form two anamorphic prism pairs and provide a beam expansion between 10× to 17×. This beam expansion enhances the grating diffraction resolution by increasing the number of grooves illuminated, as well as reducing the power density on the grating. In order to minimize the reflection loss, the right angle sides of the prisms are all antireflection (AR) coated and the diagonal sides are set close to Brewster's angle. The grating is mounted on an encoded rotary stage (Aerotech ADRS-100), which is computer controlled. The angular resolution of the rotary stage is 0.87 μrad and provides a tuning resolution of ~ 0.3 GHz around the wavelength of 800 nm. The output coupler is a flat mirror with 80±3 % reflectivity at 700-900 nm on the surface facing the cavity and AR coating on the other side. Details of this grating laser are described elsewhere [16].

In this work, a PR mirror was inserted in the grating arm of the resonator, between the curved mirror and the prism set. This creates a second, shorter F-P resonator with the output coupler. The reflectivity of the PR mirror is an important parameter in this experiment: it shall be high enough so that F-P resonator can build up power and oscillate; and it also shall distribute enough power to the grating cavity, so that the wavelength-selection feedback or seeding of the grating cavity to F-P resonator can be strong enough to suppress all adjacent longitudinal modes other than the seeded mode.

### 3.    Results and discussion

Both cavities can lase independently. Firstly we chose an 80% PR mirror (Layertec: AR: 0°, 700-900 nm < 0.25%, PR: 0°, 700-900 nm 80±3 %). Pumped with 9.85 W 532 nm green laser, the classic grating laser can emit 1.43 W narrow-linewidth (NL) laser light at 789 nm. The linewidth was measured as 3.5 GHz by a wavelength meter (HighFinesse: WS/6). The WS/6 wavelength meter has a set of Fizeau interferometers with a free spectral range down to 20-30 GHz for infrared light, and its linewidth measurement accuracy is 0.5 GHz. The F-P cavity alone gives a BB laser emission of 1.65 W, which was measured by putting an optical opaque blocker between the prism set and the PR mirror. When the grating cavity is coupled to the F-P cavity, the output laser power can significantly increase up to 2.22 W with the laser linewidth as 5.7 GHz. Fig. 2 shows the interferometric patterns of the laser measured with the wavelength meter for comparison.

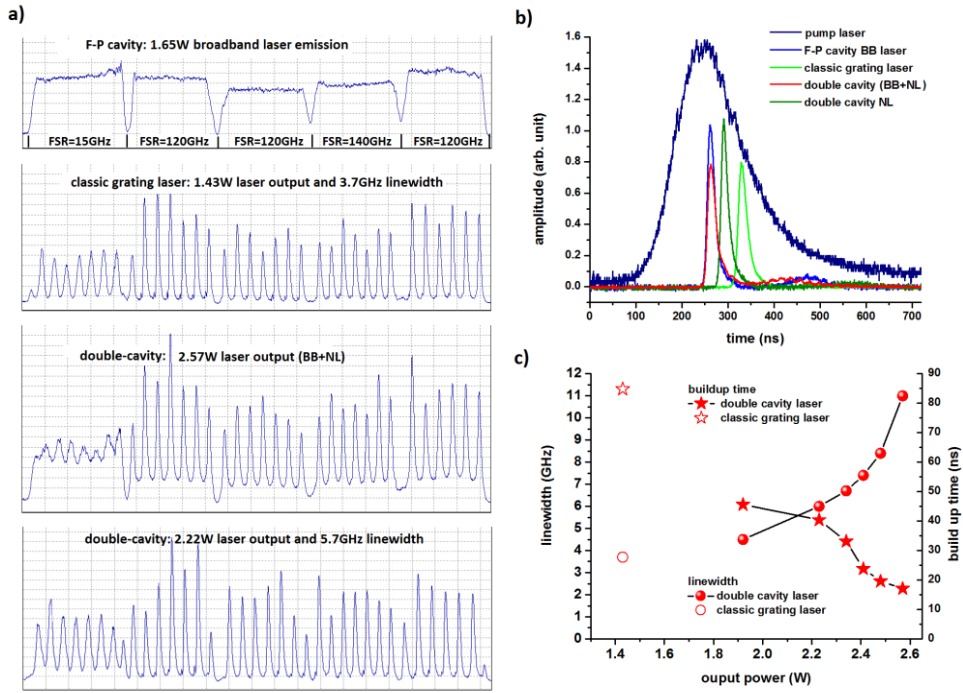

Fig. 2 The measurements were all made at 789 nm. a): interferometer patterns of the F-P cavity, the grating cavity and the double-cavity mode (using 80% PR mirror). The patterns were measured by a HighFinesse WS/6 wavelength meter. The free spectral ranges (FSR) of the five interferometers are specified in the figure. b): the pulse profile of the pump laser and Ti:Sa lasers at different configurations. c): the variations of the linewidth and the pulse buildup time with the output power for the double-cavity laser.

At the same time, the temporal profile of the laser pulses were measured with silicon photodiodes (Thorlabs, DET 10A) and a digital oscilloscope (Tektronix, DPO3014) to capture the dynamics of laser pulse formation. The results are shown in Fig. 2. Principally with the same pump power, the buildup time depends on the cavity length and cavity loss. With shorter cavity length and lower cavity loss, the F-P cavity lases first (Fig. 2).

When the pump pulse was short (tens of ns), as in Ref. 9-12, the BB mode would quickly deplete all available inversion density after the pump pulse. Therefore the NB mode wouldn't be able to oscillate. In order to avoid this effect, all the works of Ref. 9-12 chose a NB cavity with the length much shorter than that of the BB cavity. When the NB mode lases first, it will induce stimulated emission of the BB oscillator at the seeded wavelength, so that the combined system behaves like a typical injection-seeded laser.

Different from those previous works, our pump laser has a pulse width around 160 ns and the Ti:Sa laser lases typically within the pump pulse duration. Although the NB cavity (grating cavity) lases later, it still gets available population inversion from the long pump pulse. When two cavities are coupled, the double cavity can emit NL laser with the same buildup time as the F-P cavity. When the feedback is not strong enough, the double cavity can emit two components - NL and BB lasers - at the same time. This phenomenon can be observed in the laser interference pattern measured by the wavelength meter WS/6 (Fig. 2-a BB+NL pattern). The background indicates the BB component, and interference peaks indicate the NL component. However the NL-only mode can be still achieved by slightly turning the PR mirror off the cavity axis, which increases the relative feedback strength by suppressing the gain of the F-P cavity. The output laser power is a little bit lower but the frequency purification is enhanced significantly (Fig. 2).

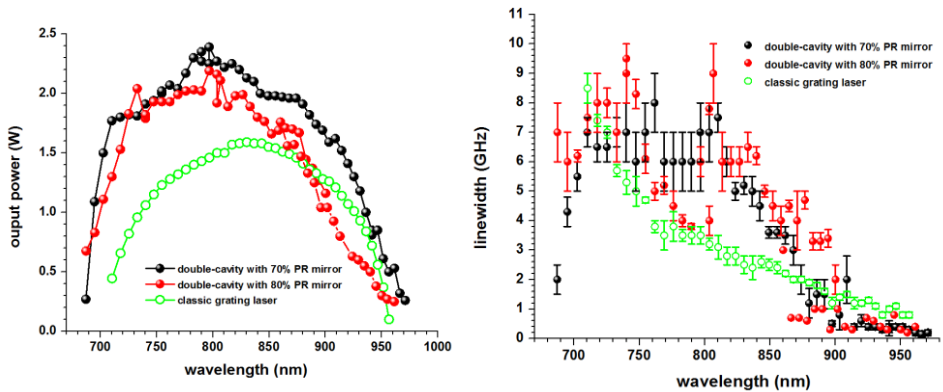

Fig. 3. The laser output powers and the linewidths at 790 nm for the classic grating laser, the double-cavity laser with the 80% PR mirror, and the double-cavity laser with the 70% PR mirror at a pump power 9.85W.

Observing insufficient feedback (especially at long wavelength end) with the 80% PR mirror, a 70% PR mirror (Layertec: AR: 0°, 700-900 nm < 0.25%, PR: 0°, 740-850 nm 70±3 %) was used to form the coupled cavities. For the complete Ti:Sa laser wavelength range (680-980 nm), the output laser powers and the corresponding linewidths for both cases were measured and are presented in Fig. 3. For comparison, the output power and laser linewidth for the classic grating laser are also plotted. The result shows the double-cavity configuration generally enhances the output laser power by a factor of 1.5-2.0 without a significant increase of the linewidth. The improvement is more pronounced on the short wavelength side. With increasing wavelength the pulse buildup time in the grating cavity increases. This can be attributed to the wavelength-dependent first-order diffraction efficiency of the grating. When the pulse buildup time runs far out of the pump pulse duration, the seeding stimulation cannot establish any more. However, a laser emission can still be observed by titling the PR mirror more off the cavity axis, which restores the system to a classic grating laser with a gain loss on the PR mirror. This mode switching happens around 880 nm for the 80% PR mirror, and around 940 nm for the 70% PR mirror, which are both observable in the power and linewidth variations in Fig. 3. Meanwhile it can be directly observed by monitoring the time profile of the laser pulse. Due to the higher gain and short cavity length, the BB mode pulse always fires before the NL mode pulse. At the moment when two modes couple to each other, the NL mode pulse is observed to vanish suddenly, or in more vividly description "merge into the BB mode pulse". However after restoring to the classic grating mode, a NL laser emits only if the BB mode pulse was delayed to overlap with the grating cavity pulse.

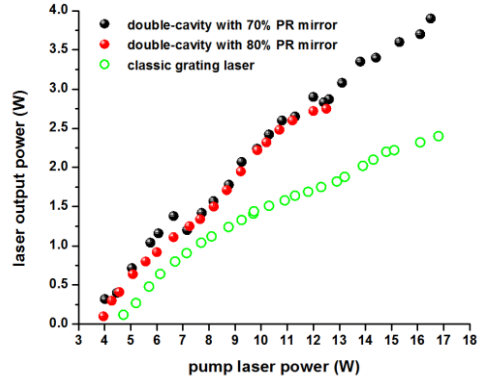

Fig. 4. Dependence of the laser output power and lasing threshold at 790 nm on the pump power for the classic grating laser, and the double-cavity lasers with a 80% PR mirror and a 70% PR mirror respectively.

Fig. 4 shows the output power of the double-cavity laser (for both 80% and 70% PR mirror cases) as a function of the pump power at 790 nm, with comparison to the classic grating laser. The double-cavity configuration lowers the threshold from 4.5 W (classic grating laser) to 3.7 W. At low pump power 4-8 W, both the double-cavity and classic grating laser show similar slope efficiency ~33%. However for higher pump power greater than 8 W, the double-cavity laser retains the same slope efficiency while the classic grating laser switches mode to lower slope efficiency ~15%. One possible explanation is a deformation of the gold coating of the grating under a high intracavity oscillation power. In the coupled resonator case, this is alleviated by confining most of the oscillating power inside the F-P cavity.

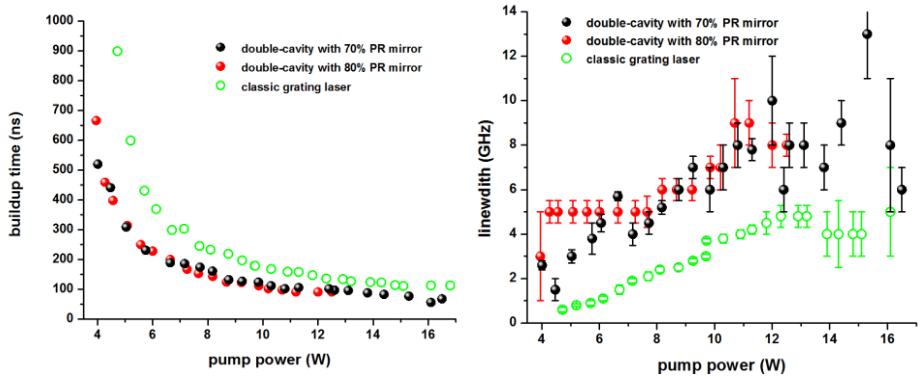

Fig. 5. Dependences of the laser pulse buildup time and the linewidth at 790 nm on the pump power for the classic grating laser and the double-cavity laser with the 80% PR and the 70% PR mirror.

The variations of the laser pulse buildup time and linewidth with the pump power have been investigated (Fig. 5). The buildup time is defined as the difference from the 25% leading edge of the pump pulse to that of the laser pulse released [11]. From the result, the buildup time considerably reduces with the increase of the pump power for all configurations. On the other hand, the linewidth generally increases with the pump power and flattens after 12 W, which corresponds to the tendency that the buildup time approaches its smallest value after 12 W Fig. 5 left). The linewidth of the double-cavity configuration generally doubles compared to the classic grating laser. The 70% and 80% PR mirror cases behave similarly at the wavelength 790 nm.

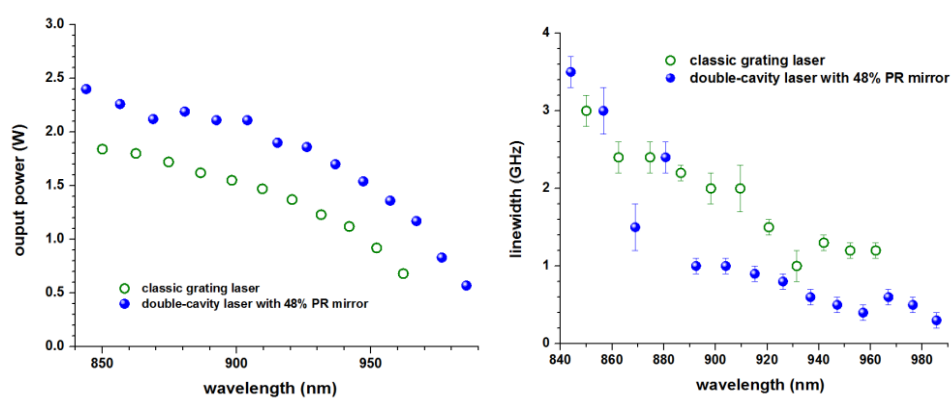

Fig. 6. The output laser power and linewidth of the double-cavity laser with the 48% PR mirror at the pump power of 11.6W. The classic grating laser performance at the same pump power is plotted for comparison.

It can be seen from Fig. 5 that the buildup time significantly decreases when increasing the pump power. For wavelengths >850 nm a higher pump power may compensate the higher grating diffraction loss, and therefore keeps the system in the double-cavity NL mode instead of switching back to the classic grating laser mode. An investigation on the double-cavity configuration (70% PR mirror) with a higher pump power (11.6 W) was attempted for long wavelength operation. However the result shows a higher pump power increases the gain of the F-P cavity more significantly, which makes the seeding power from the grating cavity not enough to suppress the broadband emission. A balance between the pump power and the PR mirror reflectivity has to be pursued. For the long wavelength operation, a 48% PR mirror (Layertec: AR: 45°, 500-980 nm < 2%, PR: 45°, 440-1040 nm 48±2%) was tested with 11.6 W pump power. The NL mode with ~ 35% power improvement was achieved compared to the classic grating laser at the same pump power.

## 4. Conclusion

In this work we reported a new pulsed double-cavity Ti:Sa laser configuration obtained by inserting a PR mirror into a classic Littrow grating laser cavity, thus creating a coupled-cavity oscillator-amplifier configuration. A 1.5 to 2 fold increase of output power with respect to the classic grating laser configuration was obtained. The linewidth of the new configuration is the same as, or 1.5-2 times higher than that of the classic grating laser depending on the operating wavelength. The detailed characteristics of the new laser configuration with 70% and 80% PR mirrors were measured. A performance improvement at the long wavelength range (> 850nm) was achieved with a 48% PR mirror at a higher pump power. The new system not only provides a continuously tunable laser source for 690-970 nm with the enhanced pulse peak power up to 5 kW and the linewidth of ~ 6 GHz, but also an ideal source for tunable second-harmonic generation into 350-480 nm laser. It will be well suited to isotope separation, resonance laser ionization spectroscopy and LiDAR. The performance of this double-cavity system could be further improved by adapting the PR mirror reflectivity to the Ti:Sa gain curve.

**Acknowledgments**

The author would like to thank the funds provide by TRIUMF which receives federal funding via a contribution agreement with the National Research Council of Canada and through a Natural Sciences and Engineering Research Council of Canada (NSERC) Discovery Grant (386343-2011). M. Mostamand acknowledges funding through University of Manitoba graduate fellowship.